\documentclass[aps,prd,floatfix,twocolumn,footinbib,altaffilletter]{revtex4-1}

\usepackage[colorlinks=true,citecolor=blue,linkcolor=blue]{hyperref}
\usepackage{amsmath,amssymb}
\usepackage{epsfig}  
\usepackage{graphicx}               
\usepackage{url}
\usepackage{hyperref}
\usepackage{color}
\usepackage{placeins}
\usepackage[update,prepend]{epstopdf}
\usepackage{soul}

\usepackage{titlesec}
\titleformat{\section}{\normalfont\bfseries}{\thesection}{1em}{} 
\titlespacing*{\section}{0pt}{8pt}{8pt}

\def\thesection{\Roman{section}}

\begin{document}

\title{Upper Limit on Forward Charm Contribution to Atmospheric Neutrino Flux}

\author{Francis Halzen\footnote{francis.halzen@icecube.wisc.edu}}
\author{Logan Wille\footnote{wille3@wisc.edu}}
\affiliation{Wisconsin IceCube Particle Astrophysics Center and Department of Physics, University of Wisconsin, Madison, WI 53706, USA} 

\begin{abstract}

We revisit the calculation of charm particle production in hadron collisions, focusing on the production of charm particles that carry a large fraction of the momentum of the incident proton. In the case of strange particles, such a component is familiar from the abundant production of $K^+\Lambda$ pairs. Modern collider experiments have no coverage in the very large rapidity region where the forward pair production dominates. While forward charm particles are produced inside the LHC beampipe, they dominate the high-energy atmospheric neutrino flux in underground experiments because long-lived pions and kaons interact before decaying into neutrinos. The fragmentation of the {\it spectator} quark in the partonic subprocesses $qc \rightarrow qc$ and $gc \rightarrow gc$ is responsible for the forward component of charm production in perturbative QCD. We use this phenomenological framework to construct a charm cross section that saturates available accelerator and cosmic ray data, i.e., it represents an upper limit on the normalization of the charm cross section that cannot be reliably calculated because the charm mass is much smaller than the center-of-mass energy. Where the highest energy IceCube observations are concerned, we conclude that the upper limit on the flux of neutrinos from forward charm production may dominate the much-studied central component. It may therefore also represent a significant contribution to the TeV atmospheric neutrino flux but cannot accommodate the PeV flux of high-energy cosmic neutrinos observed by IceCube, or even the excess of events observed in the 30 TeV energy range.
\end{abstract}

\maketitle

\section{Introduction}
\label{sec:intro}

The production of charm hadrons by cosmic rays interacting in the Earth's atmosphere \cite{Enberg:2008,Gauld:2015kvh,Gauld:2015yia,Bhattacharya:2015jpa,Garzelli:2015psa,Pasquali:1998ji, Gondolo:1995fq,Gaisser:2013rx,Gelmini:2000wm,Martin:2003bu} is the dominant background for the detection of cosmic neutrinos above an energy that depends on the charm cross section and on its dependence on Feynman $x_F$. Because of their short lifetime, charm hadrons decay promptly into neutrinos in contrast with relatively long-lived high-energy pions and kaons that, at high energies, interact and lose energy before decaying. Although prompt neutrinos may represent the dominant component of the atmospheric neutrino background for the identification of the cosmic neutrino flux at PeV energy, they have not yet been identified as such. IceCube observations \cite{JVS:2014} indicate that the neutrino flux is dominated by conventional atmospheric neutrinos at low energy and by cosmic neutrinos at high energy; charm neutrinos never dominate the measured spectrum. Obviously, the issue is of great interest because a poor understanding of a potential charm neutrino background interferes with the precise characterization of the cosmic flux measured by IceCube. Neutrinos provide the first unobstrutced view of cosmic accelerators at the highest energies.

It is important to realize that the production of charm in the atmosphere cannot accommodate the observed flux of high energy neutrinos. We know, independent of theory, that the charm flux tracks the energy dependence of the cosmic ray flux incident on the atmosphere and that it is indipendent of zenith angle. There is no evidence for such a component in any of the multiple IceCube analyses. On the other hand there is accumulating evidence for a steepening of the cosmic neutrino flux with lower threshold; the flux is not a single power. Already the first attempt to lower the threshold \cite{JVS:2014} revealed an excess of events in 30 TeV energy range raising the possibility of a charm background.

The hadronic production of charm particles has been extensively studied in the context of perturbative QCD \cite{Beenakker:1988bq,Ellis:1988sb,Nason:1987xz,Nason:1989zy}. These calculations often use a color dipole description of the target proton \cite{Arguelles:2015wba,Nikolaev:1995ty,Raufeisen:2002ka,Kopeliovich:2002yv} in order to mitigate the breakdown of the perturbative calculation associated with large $log(1/x)$ where $x=m_c / \sqrt{s}$. Here $m_c$ is the charm quark mass and $s$ the center-of-mass energy of the colliding hadrons. At high energy, the charm quark is no longer a heavy quark whose mass controls the perturbative expansion. More importantly, these calculations only describe the central production of charm particles with a cross section that peaks at Feynman $x_F \sim 0$, providing an incomplete picture. For strange particles, the central component of particle production is accompanied by a forward component where the incident proton transfers most of its energy to a $K^+ \Lambda$ pair with the same quantum numbers \cite{Edwards:1978mc}. It dominates strange particle production at large Feynman $x_F$. In this paper, we evaluate the corresponding cross section for charm production in perturbative QCD making no attempt to compute its normalization, which is highly uncertain as it is for the central component. However, limits on its magnitude can be derived using IceCube observations of the atmospheric flux of muon and electron neutrinos and data from archival experiments performed at the CERN Intersecting Storage Ring (ISR). We will thus obtain a ``maximal'' contribution of the forward component of charm particle production that, like for strange particles, contributes qualitatively at the same level as the central component to the total charm particle cross section. However, while it dictates the production of the highest energy atmospheric neutrinos in IceCube, we conclude that it cannot accommodate the flux of cosmic neutrinos that dominates the spectrum at the highest neutrino energies. In addition, this forward charm production is unable to explain the 30 TeV excess over the best-fit power law  as seen in a recent IceCube analysis focusing on lower energy neutrinos \cite{JVS:2014}.

\begin{figure}[h]
\includegraphics[width=0.48\textwidth]{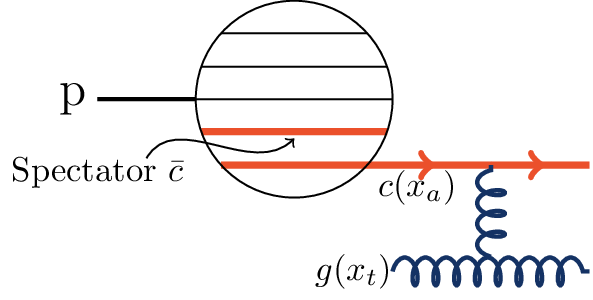}
\caption{A partonic interaction between two protons via a c-$\bar{c}$ quark pair in one proton and a gluon in the other. We refer to the charm quark that interacts with the gluon as ``active,'' the other as ``spectator.''}
\label{fig:spec}
\end{figure}

It has been pointed out some time ago that a perturbative QCD calculation of charm produces a forward component that peaks near $x_F \sim 1$ where it dominates the central component \cite{Halzen:Diff}. The forward charm particles are the hadronization product of the spectator quark in the leading-order partonic subprocesses $qc \rightarrow qc$ and $gc \rightarrow gc$; see Fig. \ref{fig:spec}. While the active constituent charm quark carries a momentum fraction $x \sim0$, the spectator quark hadronizes with the valence quarks of the incoming proton into a charm hadron that carries a large fraction of the proton momentum. The high-momentum charm hadrons thus produced decay leptonically to energetic muons and neutrinos that penetrate into underground detectors. The spectrum of these neutrinos extends higher in energy than predicted by calculations that have neglected the forward component. While this prompt neutrino flux cannot explain the high-energy neutrino flux observed by IceCube, it potentially represents a background to the cosmic neutrino flux and is therefore critical to characterize.

This paper is organized as follows: in section \ref{sec:diff}, the differential cross section for production of the spectator charm quark is derived within the conventional framework of perturbative QCD. In section \ref{sec:flux}, we evaluate the associated upper limit flux of prompt neutrinos. We subsequently discuss in section \ref{sec:veto} the effect of the IceCube veto routinely imposed in IceCube analyses. It requires that candidate cosmic neutrinos are not accompanied by particles that signal the presence of an air shower. We finally confront our estimates of the charm flux with the data in section \ref{sec:conclu}.

\section{The Forward Charm Cross Section}
\label{sec:diff}
We first perform the leading order calculation of charm quark production including the diagrams involving constituent charm quarks in the interacting hadrons, 

\begin{equation}
\frac{\partial \sigma(pp \rightarrow \bar{c}_a)}{\partial x_a} = \sum_f \int  \,dx_t\, c(x_a,Q^2)\, f(x_t,Q^2)\, \hat{\sigma}_{f \,\bar{c} \rightarrow f \,\bar{c}},
\label{eq:dxF}
\end{equation}
where the incoming proton interacts via its charm constituents described by the parton distribution function (PDF) $c(x_a,Q^2)$. Here, $f(x_t,Q^2)$ is the PDF of the target, $x_a$ is the active charm quark fractional momentum, $x_t$ is the target parton fractional momentum and $\hat{\sigma}_{f \,\bar{c} \rightarrow f \,\bar{c}}$ is the partonic cross section. We sum over the diagrams involving a charm constituent of the interacting hadrons shown in Fig. \ref{fig:feyn}. All subprocesses are explicitly given in the appendix. We fix the two free parameters, $Q^2$ and $\hat{t}_{min}$, to $3~m_c^2$ and $2~m_c^2$, respectively, to match $c \bar{c}$ cross section data following references~\cite{Alice:2012,ATLAS:2011fea,Aaij:2013mga,Adare:2006hc,Adamczyk:2012af}. Whereas the charm structure function of the proton could at best be guessed at in reference~\cite{Halzen:Diff}, the charm partonic distribution function has now been determined by collider experiments. For this paper, we use the PDF set CT10 \cite{Lai:2010vv}.

We next focus on the contribution of the charm cross section where the spectator quark bleaches its color by combining with the valence quarks in the incident proton,

\begin{equation}
\begin{aligned}
\frac{\partial \sigma(pp \rightarrow c_s)}{\partial x_s} =& \int\ dx_{u_1} dx_{u_2} dx_d dx_a \ \bigg[\frac{\partial \sigma(pp \rightarrow \bar{c}_a)}{\partial x_a} \\
&P(x_{u_1},x_{u_2},x_d) \delta(1-x_s-x_a-x_{u_1}-x_{u_2}-x_d) \bigg],
\label{eq:dxL}
\end{aligned}
\end{equation}
where $x_s$ is the spectator charm quark fractional momentum and $x_{u1}, x_{u2}, x_d$ are the valence quarks fractional momenta. We enforce $x_s = 1 - x_a - x_{u_1} - x_{u_2} - x_d$ such that the spectator charm quark carries all momentum not carried by the valence quarks or active charm quark. Finally, $P(x_{u_1},x_{u_2},x_d)$ is the probability distribution of the valence quarks fractional momentums given by the normalized parton distribution functions of the valence quarks,

\begin{equation}
P(x_{u_1},x_{u_2},x_d) = \frac{u_v(x_{u_1})}{2} \frac{u_v(x_{u_2})}{2} d_v(x_d). 
\label{eq:P}
\end{equation}

\begin{figure}[h]
\includegraphics[width=0.48\textwidth]{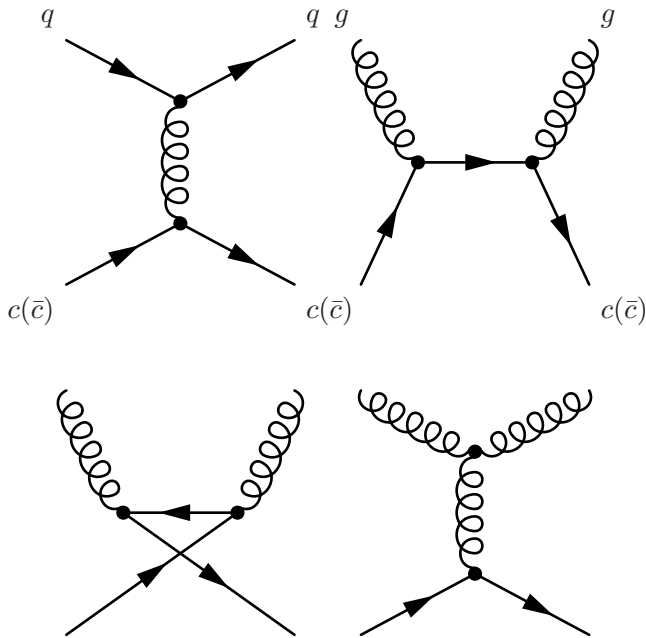}
\caption{The partonic interaction diagrams that lead to the production of a spectator charm quark in a proton-proton interaction.}
\label{fig:feyn}
\end{figure}

\begin{figure}[h]
\includegraphics[width=0.5\textwidth]{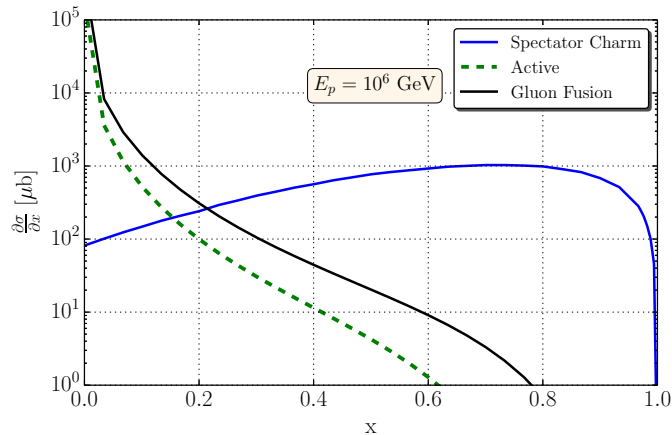}
\caption{The differential cross sections for producing charm quarks as a function of their longitudinal momentum fraction. The central gluon fusion component is shown along with the contribution of the diagrams with an active charm quark. The spectator charm's differential cross section is peaked at large $x_F$ carrying the majority of the momentum of the proton that produced it.}
\label{fig:comparexF}
\end{figure}

The resulting charm  cross section is shown in Fig. \ref{fig:comparexF} for an incident proton energy $E_p=10^6$ GeV. The forward component is shown along with the central component dominated by the subprocess $gg \rightarrow c \bar{c}$, Eq. (\ref{eq:gg}). While the active charm component represents a subdominant contribution to the central production of charmed particles, the spectator contribution dominates for x-values above 0.2. The spectator charm effectively becomes a valence quark during this interaction, resulting in the forward production. Note that perturbative QCD  generates a forward component of the charm cross section without resorting to intrinsic charm. The subprocess effectively promotes the spectator charm to a valence quark. The phenomenology is therefore similar to that of intrinsic charm quarks in the proton and we do not anticipate that an analysis modeling intrinsic charm will lead to different conclusions.

\begin{figure}
\includegraphics[width=0.5\textwidth]{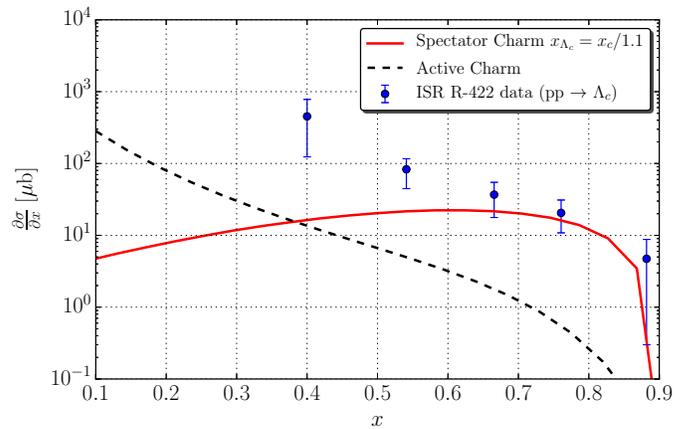}
\caption{The Feynman $x_F$ dependence for $\Lambda_c$ production is compared with ISR data \cite{Bari:1991in} at $\sqrt{s} = 63$ GeV.}
\label{fig:Hadron}
\end{figure}

A caveat at this point is that the spectator charm cross section calculates the production of a charm quark and not a charmed hadron. The hadronization of the spectator charm quark with the valence quarks of the incident proton results in the dominant contribution to forward scattering. The strings linking the charm quark with the other proton constituents into charm hadrons reduce its fractional momentum. We implement a hadronization scheme inspired by reference~\cite{Martin:2003us} by reducing the charmed baryon momentum, $x_{\Lambda_c} = {x_s}/{1.1}$. In Fig. \ref{fig:Hadron}, we show that this procedure actually describes the $x_F$ dependence of the highest energy measurement available from ISR R-422. This archival data represent the strongest constraint on how much momentum can be carried by $\Lambda_cD$ pairs.

The calculation described above leads to atmospheric neutrino fluxes that exceed the data. As the normalization of the cross section is sensitive to the scales defining the perturbative QCD expansion, we let it float and obtain a maximal value for a normalization $\sigma(\Lambda_c) / \sigma(c_s) = 0.214$. In other words, there is a tension in normalization between the atmospheric neutrino and ISR data; see also Section \ref{sec:flux}. We repeat the same procedure for $\bar{D}^0$ and $D^{-}$ meson hadronization with a shape change of $x_{D} = {3 x_s}/{4}$ and cross section normalization of 0.476 and 0.238 respectively.

An important note, the hadronization scheme chosen is maximal without exceeding measured neutrino fluxes. We thus obtain an upper limit of the forward charm contribution to the neutrino flux and explore its potential impact on the IceCube events observed. Although it is straightforward to include the higher order diagrams, the changes introduced are well within the uncertainties associated with the hadronization that is poorly constrained by data. 

\section{Prompt Neutrino Flux Upper Limit}
\label{sec:flux}
We next calculate the neutrino spectrum from the decay of the charmed hadrons using the MCEq atmospheric interaction package \cite{Fedynitch:2015zma}. We use two parameterizations of the incident cosmic ray flux from Gaisser and collaborators \cite{ref:Id0,Gaisser:2012,Gaisser:2013}. These fits to the cosmic ray flux assume a very different primary composition, yet they yield very similar results in this context. We use the H3a fit, which assumes a heavy nuclei composition at high energies, for the rest of the paper.

\begin{figure}[ht!]
\includegraphics[width=0.5\textwidth]{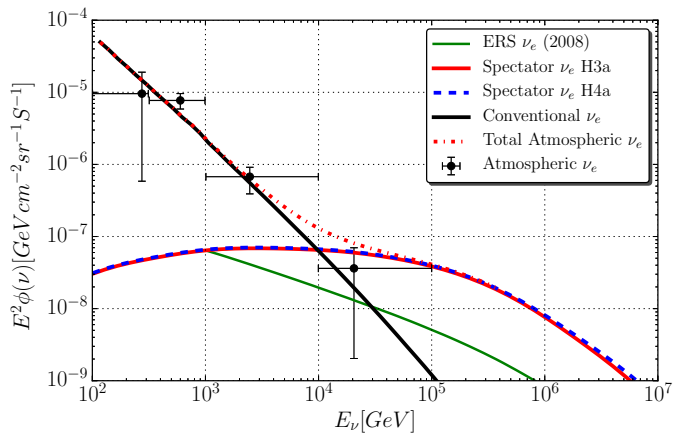}
\caption{The maximal prompt neutrino spectrum for the spectator charm and Enberg et al. calculations. The spectator $\nu_e$ flux saturates the 10 TeV atmospheric $\nu_e$ flux measurement, showing the normalization for the charm meson hadronization cannot be larger. Note, the ERS flux is only shown down to 1 TeV.}
\label{fig:flux}
\end{figure}

The neutrino spectrum upper limit resulting from the maximal contribution of the hadronization of the spectator charm described in the previous section is compared to the highest energy measurements \cite{Aartsen:2015xup} of the atmospheric electron flux in Fig.~\ref{fig:flux}. It is compared to the one of Enberg et al. referred to as ERS \cite{Enberg:2008}. Note that our floating normalization spectator charm neutrino flux added to the conventional pion and kaon neutrino flux \cite{Honda:2006qj} saturates the atmospheric electron neutrino flux measured by IceCube \cite{Aartsen:2015xup}. A future measurement with higher statistics will be very useful in this context. IceCube has also performed an analysis \cite{JVS:2014} of neutrino events starting in the detector that has resulted in an upper limit of the prompt neutrino flux. We note however that the flux was modeled after the ERS flux and only the normalization was varied. As seen in Fig.~\ref{fig:flux}, the spectator charm neutrino spectrum has a different shape, closer to an $E^{-2}$ spectrum below 100 TeV. This allows the model to partially evade the current IceCube limits.

\begin{figure}[h!]
\includegraphics[width=0.5\textwidth]{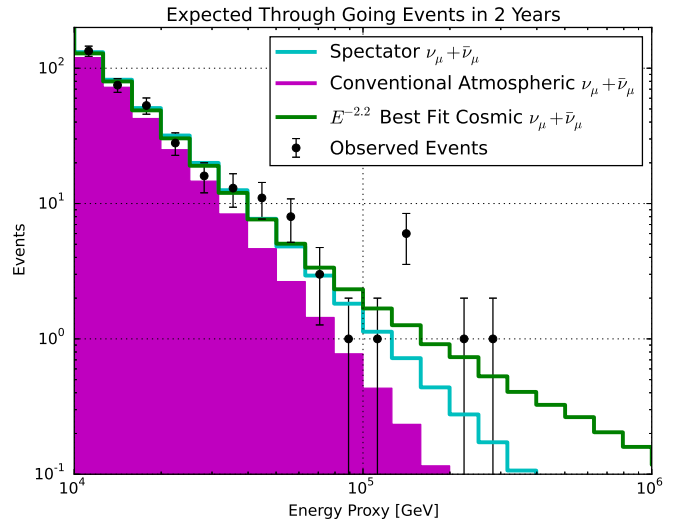}
\caption{The expected number of events in two years of IceCube looking for $\nu_\mu$ events coming through the Earth. We compare the spectator neutrino expected number of events to the best fit astrophysical spectrum found in \cite{Aartsen:2015rwa}.}
\label{fig:numuevents}
\end{figure}

We also calculated the expected number of $\nu_\mu$ events penetrating the Earth; the result is compared to the first two years of IceCube \cite{Aartsen:2015rwa} in Fig. \ref{fig:numuevents}. It is also compared to the best fit cosmic neutrino spectrum. An important point about the flux from the spectator charm neutrino is the flavor and neutrino antineutrino ratio: for the astrophysical neutrino flux the flavor ratio is assumed to be 1:1:1 for $\nu_e : \nu_\mu : \nu_\tau$ and equal parts neutrino and antineutrino. In contrast, for the spectator neutrino flux the flavor ratio is $\sim$ 1:1:0 and the neutrino to antineutrino ratio is $\sim$ 1:10. This is important when comparing the spectator neutrino flux for different analyses as they tend to prefer specific neutrino flavors.

While one may be tempted to conclude that the spectator neutrino flux may accommodate the data, this is not the case. The updated analysis using six years of IceCube livetime has revealed a high-energy astrophysical spectrum of $E^{-1.91}$, including a very high energy neutrino event with deposited energy of 2.6 PeV \cite{Aartsen:2015zva}. The flux of charm origin shown in Fig.~\ref{fig:numuevents} cannot describe this result.

\section{Comparison to IceCube Observations}
\label{sec:veto}

We have fixed a normalization of our maximal prompt flux that saturates the IceCube measurements of the atmospheric electron neutrino flux in the ten TeV energy range as well as the muon neutrino data in the 100~TeV energy range; see Fig. \ref{fig:flux} and Fig. \ref{fig:numuevents}. Atmospheric neutrinos are produced in air showers and are consequently accompanied underground by high-energy muons produced in the same shower. Therefore, IceCube's veto based searches for cosmic neutrinos routinely introduce a so-called self-veto, where an atmospheric neutrino is vetoed when accompanied by atmospheric muons \cite{PhysRevD.90.023009}. We use the technique of reference \cite{PhysRevD.90.023009} to calculate the self-veto probability of the spectator charm neutrino flux modulo a modification that we discuss next.

\begin{figure}[h!]
\includegraphics[width=0.5\textwidth]{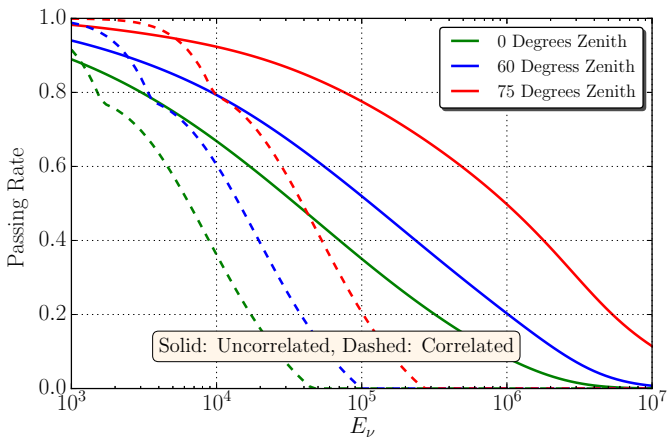}
\caption{The uncorrelated and correlated self-veto passing rate for the spectator charm neutrino flux in IceCube at three different zenith angles. The uncorrelated passing rate enforces shower energy conservation as described in the text.}
\label{fig:passing}
\end{figure}

For spectator charm, the charmed hadron carries a large fraction of the incident proton energy. This reduces the energy of the remaining shower, which reduces the probability for producing an high-energy muon in a second, uncorrelated hadron. We contrast correlated and uncorrelated muons, with the correlated produced in the hadron decay along with the observed neutrino, and the uncorrelated originating from other particles in the air shower. While negligible for central charm production, for spectator charm this effect can be significant in calculating the uncorrelated muon self-veto probability. For the correlated muon, a muon that is produced by the same hadron decay as the neutrino, there is no need to make this modification. The uncorrelated muon and correlated muon self-veto passing rates are shown in Fig.~\ref{fig:passing}. 

\begin{figure}
\includegraphics[width=0.5\textwidth]{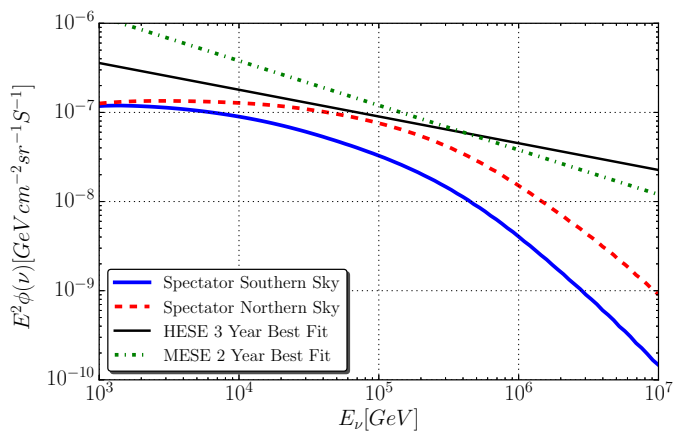}
\caption{The spectator charm neutrino spectrum summed over neutrino flavors is shown for the northern and southern sky, with the self-veto effect added to the southern sky flux. Two IceCube veto-based-analysis best fit spectra summed over neutrino flavor are shown as a reference \cite{JVS:2014, PhysRevLett.113.101101}.}
\label{fig:vetoflux}
\end{figure}

\begin{figure*}
\begin{center}
\includegraphics[width=0.95\textwidth]{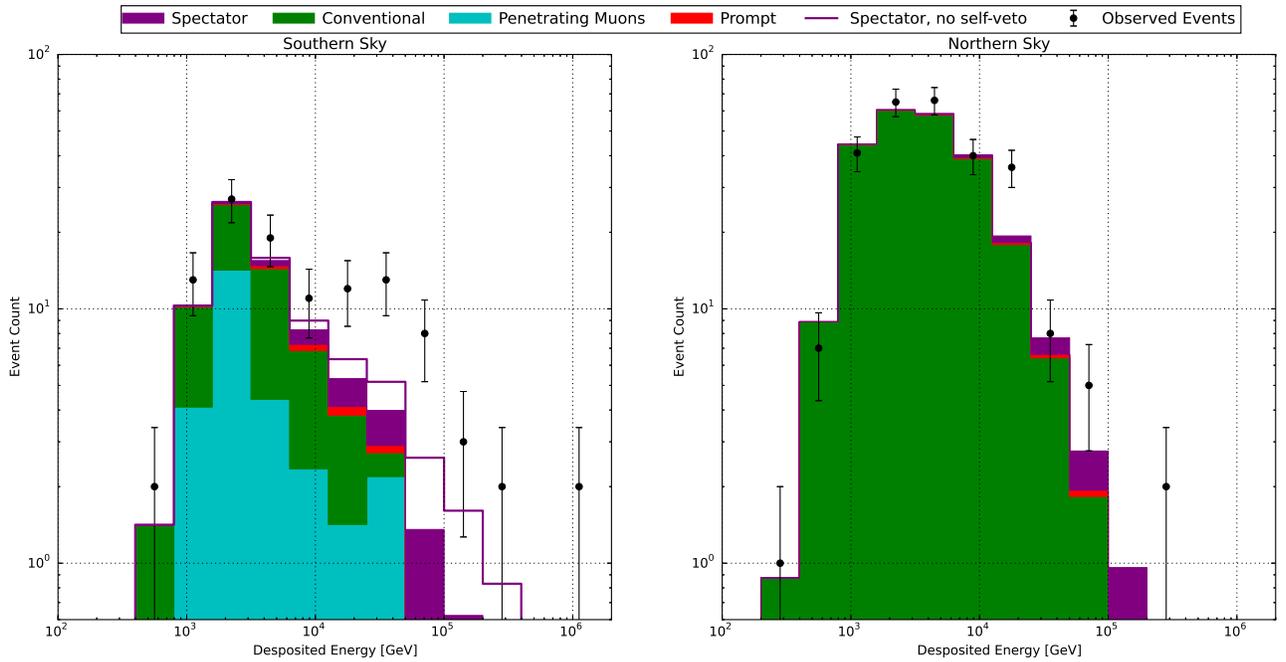}
\caption{The expected number of events in both the northern and southern sky for two years in IceCube using a veto-based detection scheme \cite{JVS:2014}. In the northern sky, the maximal flux from the spectator charm neutrino leaves little room for an additional cosmic neutrino flux without exceeding the observed events. While in the southern sky, an additional cosmic neutrino flux is needed to have agreement between the expected number of events and the observed number of events. Even removing the self-veto effect discussed in section \ref{sec:veto}, the maximal prompt neutrino flux cannot explain the high-energy events observed in IceCube.}
\label{fig:Event}
\end{center}
\end{figure*}

Using these results, we obtain Fig. \ref{fig:vetoflux}, the maximal prompt neutrino spectrum resulting from forward charm along the best fit of IceCube neutrino spectra obtained in veto-imposed analyses. We subsequently compute the corresponding number of events in the 2-year ``MESE'' IceCube analysis \cite{JVS:2014} separately for the Southern and Northern Hemispheres; see Fig. \ref{fig:Event}. For the Northern Hemisphere, it may be tempting to conclude that the data can be described by charm. This conclusion is helped by the fact that cosmic neutrinos with PeV energy and above are in any case absorbed by the Earth. However, in the southern sky, there is a significant disagreement between the observed events and the expected number of events for the extreme spectator neutrino flux that we have constructed, both where its normalization and its Feynman $x_F$ distribution are concerned. Clearly an additional astrophysical flux is required to achieve agreement between the expected and observed events. Even after removing the self-veto of the spectator neutrino flux we reach the same conclusion. The basic reason for the disagreement is the softening of the atmospheric neutrino spectrum, which is strongly suppressed above 100 TeV---the prompt flux simply traces the atmospheric cosmic ray spectrum and cannot accommodate the highest energy events.

\section{Conclusions}
\label{sec:conclu}
We have used leading-order perturbative QCD to emphasize the fact that one expects a central as well as a forward component of charm production, a fact clearly underscored by the data for strange particle production. We have computed the charm cross section and the Feynman-$x_F$ distribution of the secondary charm particles including both components with the normalization is treated as a parameter. It cannot be reliably predicted because of large logartihms associated with the small value of $m_c/\sqrt{s}$ at these energies. The normalization was maximized without exceeding the charm measurements at colliders and the high-energy atmospheric neutrino data in underground experiments. We subsequently calculated the upper limit flux of prompt neutrinos from the decay of charmed particles in IceCube, which is clearly dominated by a potential forward component of the flux. Finally, we applied the effect of self-veto on the prompt neutrino flux and showed the expected event distribution for two years of IceCube. We found that the prompt neutrino flux from a forward charm may represent a significant background to the cosmic neutrino flux but cannot explain the high-energy events observed by IceCube at the highest energies.

\section{Acknowledgments}

Discussion with collaborators inside and outside the IceCube Collaboration, too many to be listed, have greatly shaped this presentation. We acknowledge the comments of M. Reno and M. Ahlers on an earlier version of the manuscript. We also thank A. Fedynitch and J. van Santen  for assistance in modifying their programs. This research was supported in part by the U.S. National Science Foundation under Grants No.~ANT-0937462 and PHY-1306958 and by the University of Wisconsin Research Committee with funds granted by the Wisconsin Alumni Research Foundation.

\section{Appendix}
\label{sec:appen}
Here we show the partonic cross sections used in this paper with their respective threshold center of mass energy. These cross sections correspond to the Feynman diagrams shown in Fig.\ref{fig:feyn}. The cross sections follow in Eqns. \ref{eq:qc} through \ref{eq:gg}, with kinematic variables $t_{min} = 2 {m_c}^2$, $y_0 = (\hat{s}^2 - m^2)^2/\hat{s}$, $\hat{t}_0 = min (\hat{s} - m^2 - \hat{t}_{min}, y_0)$, $x_0 = (1 - 4 m_c^2/\hat{s})^{1/2}$ and $\sigma_0 = 4 \pi \alpha_s^2(\mu_F)/(3 \hat{s})$.

\begin{widetext}
\begin{equation}
\hat{\sigma}(qc \rightarrow  qc) =  \frac{\sigma_0}{3}\bigg[\bigg( 1 - \frac{\hat{t}_{min}}{y_0}\bigg) \bigg( 1 +\frac{2\hat{s}}{\hat{t}_{min}} \bigg) -\frac{2 \hat{s}}{y_0} ln\bigg(\frac{y_0}{\hat{t}_{min}}\bigg)\bigg], \hat{s}_{th}=m_c^2+\hat{t}_{min}/2+(m_c^2\hat{t}_{min}+\hat{t}_{min}^2/4)^{1/2}
\label{eq:qc}
\end{equation}

\begin{equation}
\begin{aligned}
\hat{\sigma}(gc \rightarrow  gc) &=  \frac{3\sigma_0}{4y_0}\bigg[\bigg[ 1 + \frac{4\hat{s}}{9 y_0} \bigg( 1 +\frac{m_c^2}{\hat{s}} \bigg)^2  - \frac{2(\hat{t}_0+\hat{t}_{min}}{9 ( \hat{s}-m_c^2)}+\frac{2\hat{s}y_0}{\hat{t}_0\hat{t}_{min}}+\frac{16 m_c^4}{9(\hat{s}-m_c^2-\hat{t}_0)(\hat{s}-m_c^2-\hat{t}_{min})}\bigg](\hat{t}_0-\hat{t}_{min})\\
&+2(\hat{s}+m_c^2) ln \frac{\hat{t}_{min}}{\hat{t}_0}+\frac{4 (\hat{s}^2-6m_c^2\hat{s}+6m_c^4}{9(\hat{s}-m_c^2)}ln\frac{\hat{s}-m_c^2-\hat{t}_{min}}{\hat{s}-m^2-\hat{t}_0}\bigg], \hat{s}_{th}=m_c^2+2\hat{t}_{min}
\label{eq:gc}
\end{aligned}
\end{equation}

\begin{equation}
\begin{aligned}
\hat{\sigma}(gg \rightarrow  c\bar{c}) =  \frac{\sigma_0}{4}\bigg[\bigg( 1 + \frac{4 m_c^2}{\hat{s}} + \frac{4 m_c^2}{\hat{s}^2}\bigg) ln\bigg(\frac{1+x_0}{1-x_0} \bigg) - \frac{x_0}{16}\bigg(7+\frac{31 m_c^2}{\hat{s}}\bigg)\bigg], \hat{s}_{th}=4m_c^2
\label{eq:gg}
\end{aligned}
\end{equation}
\end{widetext}




\clearpage

\bibliographystyle{apsrev}
\bibliography{Spectator}

\end{document}